\documentclass[11pt]{article}
\usepackage{graphicx}
\usepackage{pdflscape}
\usepackage{afterpage}
\usepackage{authblk}
\usepackage[margin=1.25in]{geometry}
\usepackage[usenames,dvipsnames]{color}
\usepackage{url}
\usepackage[colorlinks = true,
            linkcolor = blue,
            urlcolor  = blue,
            citecolor = blue,
            anchorcolor = blue]{hyperref}

\newcommand\snowmass{\begin{center}
\rule[-0.2in]{\hsize}{0.01in}\\%\rule{\hsize}{0.01in}\\ 
\vskip 0.1in 
Submitted to the Proceedings of \\ the U.S. Particle Physics Community Planning Exercise (Snowmass 2021)\\ 
\rule{\hsize}{0.01in}\\%\rule[+0.2in]{\hsize}{0.01in}
\end{center}}

\begin{document}

%\pubblock

\title{\textbf{Test Beam and Irradiation Facilities}}

\author[1]{M.~Hartz}
\author[2]{P.~Merkel\footnote{corresponding author, email: petra@fnal.gov}} \author[2]{E.~Niner}
\author[3]{E.~Prebys}
\author[4]{N.~Toro}
\affil[1]{TRIUMF, Vancouver, British Columbia, Canada}
\affil[2]{Fermi National Accelerator Laboratory, Batavia, 
Illinois 60510, USA}
\affil[3]{University of California Davis, Davis, California 95616-8677, USA}
\affil[4]{SLAC National Accelerator Laboratory, Menlo Park, California 94025, USA}

\maketitle

\snowmass

\abstract{
   Progress in particle physics depends on a multitude of unique facilities and capabilities that enable to advance detector technologies. Among others, key facilities involve test beams and irradiation facilities, which allow users to test the performance and lifetime of their detectors under realistic conditions. Test beam facilities are particularly important for collider and neutrino detector applications, while irradiation facilities are crucial for collider as well as some space-based astro particle detectors. This contributed Snowmass paper aims to summarize existing test beam and irradiation facilities as well as develop the need and proposals for future facilities. 
}

\maketitle

%\snowmass

\newpage 

\section{Executive Summary}

Progress in particle physics depends on a multitude of unique facilities and capabilities that enable the advancement of detector technologies. Among these are test beams and irradiation facilities, which allow users to test the performance and lifetime of their detectors under realistic conditions. Test beam facilities are particularly important for collider and neutrino detector applications, while irradiation facilities are crucial for collider as well as some space-based astro particle detectors. In the area of irradiation facilities overlapping needs between detector and accelerator instrumentation as well as targetry development can be addressed. 

As summarized in the recent report on Basic Research Needs for High Energy Physics Detector Research and Development~\cite{brn2019}, future particle accelerator-based experiments will experience unprecedented radiation exposures of 10 GigaGray ionizing dose and fluences of 10$^{18}$ 1 MeV n$_{eq}$/cm$^2$ over their lifetimes. Detectors, support structures, electronics, data transmission components, and on-board data processing units will all need to be evaluated for performance at these dose rates and integrated doses two orders of magnitude greater than systems operating today. This will require new high-dose-rate environments for accelerated testing so that the anticipated integrated dose can be delivered in days rather than months or years. Activation of the detector materials will limit facility throughput unless remote material handling techniques are implemented. Such techniques are already employed for irradiation and post irradiation evaluation of accelerator components and for targets and windows for neutrino beam lines.

A broad array of test beam capabilities are also required to develop the next generation of detectors for tracking, calorimetry, and particle identification.  In particular, access to beams of different particle species at a wide range of energies is critical to test the response and performance of different detector technologies, where well-calibrated instrumentation such as trigger hodoscopes, tracking telescopes, and Cherenkov detectors is often required to analyze and understand the performance of the device under test.  Meanwhile, some of the capabilities being pursued for next-generation detectors place stringent demands on test beam time structure. For example, testing detector performance at high event rate and extreme (ps-scale) time resolution require test beams with high repetition rate and extremely short pulses, respectively.  The BRN report \cite{brn2019} calls for ``a significant revitalization of U.S. [test beam] facilities...to enable the detailed tests required'' and highlights very precise test beams (in multiple dimensions ---  spatial, energy spread, timing, and intensity) as an important area where U.S.-based facilities can play a world-leading role.   

Meeting the needs of this diverse community motivates the support of a variety of test beams, with different features and capabilities, across the DOE complex as well as the use of international facilities.  This whitepaper surveys test-beam needs within the HEP instrumentation community and describes several opportunities for  new facilities, improvements to existing  test-beams, and detector test platforms that will provide essential and mutually complementary support for instrumentation development. Prospects discussed include: plans and applications for a high-intensity proton irradiation facility at Fermilab; a new facility with improved infrastructure for test beam experiments at Fermilab (FTBF); a new multi-purpose electron beamline at SLAC capable of delivering high repetition rates and short pulses with precise, flexible timing for test beam applications; and the Water Cherenkov Test Experiment (WCTE) at CERN. 

\section{Community Needs}

\subsection{Community Survey}

\begin{figure}
 \centering
  \includegraphics[width=0.8\linewidth]{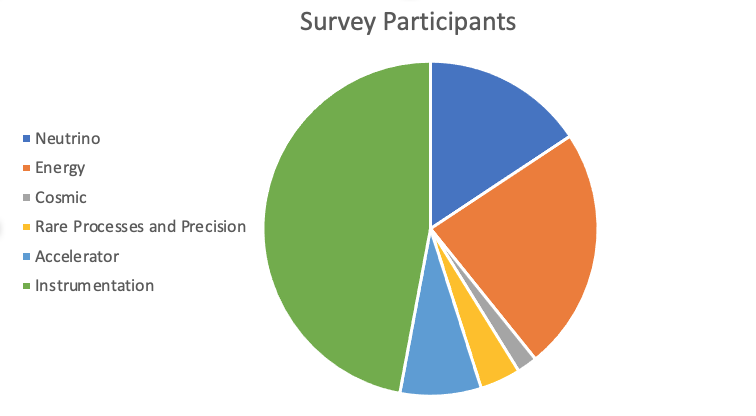}
 \caption{Distribution of research interests among the 33 participants in the survey on test beam facility needs.}
 \label{fig:survey}
\end{figure}

A survey was sent out to the test beam and irradiation facility users to solicit community feedback.  The survey has some specific focus on future needs of the Fermilab Test Beam facility but has been broadened to the general community.  This report compiles feedback from 33 participants distributed across frontiers in Figure \ref{fig:survey}.  It is still possible to participate in the survey through ~\cite{Survey}.

The test beam facilities attract a broad user base across frontiers.  There is continuing demand to use these facilities to test components for the HL-LHC upgrade.  Desired features for this work are (O)MHz proton beam with MIP energies and high quality tracking telescopes.  There is a benefit in facilities that offer both irradiations and test beam in the same location, shortening the loop in studying irradiated sensors in a test beam.  The neutrino community is makings use of test beams in the 2020s doing R\&D for DUNE, Hyper-Kamiokande and hadron production measurements with target materials.  The EIC project has steady needs over the next decade.  There is interest in beam energies from 120 GeV down to 100 MeV in all particle species.  Test beam facilities should be versatile offering as much variety in beam shape, intensity and composition as possible.  In all cases there should be robust instrumentation to finely control and characterize the beam.  The community recognizes the valuable resource test beams present for early career members and students to operate experiments.  Detector schools such as EDIT should continue to be offered. 

Long-baseline neutrino experiments such as DUNE and Hyper-K use test beams during the detector R\&D and prototyping phase in order to establish the performance of new detector technologies and techniques.  The test beams are also used to measure the response of detectors to particles of interest, providing data for detector calibration.  Neutrino experiments are generally interested in charged particles and neutrons with momenta ranging from $\sim$100~MeV/c to a few GeV/c.  Accelerator-based neutrino experiments also benefit from hadron production experiments that provide data necessary for neutrino flux calculations.  These experiments use test beam or similar facilities and may involve long-term installations, such as NA61/SHINE or short to medium-term installations such as EMPHATIC.  The beam energies of interest range from 1~GeV to 120~GeV, and the particles of interest are primarily protons, pions and kaons.

The development of novel detector technologies also places new demands on test beam. Several important new directions in detector R\&D push detector capabilities in the time domain: development of picosecond timing photodetectors; improving time resolution and response time in calorimeters; and developing tracking detectors with precise per-pixel time resolution were all highlighted as Priority Research Directions in \cite{brn2019}.  Detectors with 50 ps time resolution are planned for the HL-LHC, while future collider detectors will require 1-5 picosecond (ps) resolution.  When combined with traditional spatial measurements, such time resolutions would also enable 4D tracking with further benefits for time-of-flight-based particle identification and long-lived particle searches.  Already at the present state of the art, testing the timing precision of these detectors is challenging because it requires some external time reference of comparable or better precision to the detector being tested. In a test beam with long, continuous spills, this can only be done by putting two devices into the same beam, and correlating their responses to each other. However, this comparison conflates raw time-measurement precision with other effects, such as delays between different spatial regions of the detector.  A beam with very short pulses and an adaptable repetition rate, where the accelerator RF provides an extremely precise time reference, would be highly advantageous for understanding timing detectors’ performance. For example, a precise reference signal makes it straightforward to distinguish a distribution of timing delays across a detector (a calibration issue) from jitter or statistical variability in the response time. The critical capabilities required are (1) pulses of ps length or less, (2) the availability of a clean time reference from the accelerator and (3) the ability to set the repetition rate low enough to disambiguate the effects of different beam pulses.  At the same time, the ability to deliver a high (and controlled) repetition rate is valuable for testing detectors' responses to the high event rates expected in many  next-generation experiments. 

The Fermilab test beam facility is heavily subscribed and would benefit from modernization.  Particularly in the neutrino, muon, and dark matter communities there is interest in clean low energy beam lines with the ability to produce electrons and muons.  The beam line itself is a mile stub from the Main Injector and is facing significant aging effects and loss of spares that reduces uptime.  A new or upgraded facility should be strongly considered.  Users have expressed interest in spectrometer magnets and large aperture magnets to magnetize detectors.  An improved particle identification  system based on LAPPDs is of interest along with a unified DAQ.  An expansion in beam lines would allow the facility to better serve the combination of 120 GeV proton users and lower energy mixed species users at the same time.  With the built in facility staff and beam infrastructure the test beam also presents an ideal location for fully fledged experiments (e.g LArIAT, EMPHATIC) that do not have resources to build a new beam line but operate on a longer timescale (months to year) than typical test beams.

Irradiation areas around the world are in high demand and difficult to obtain beam time at.  Component testing for the HL-LHC upgrade is a heavy user of these facilities.  There is present demand to measure total integrated dose on the order of $10^{16}$ to $10^{17}$ 1 MeV n$_{eq}$/cm$^2$ in a reasonable timescale.  Future collider needs would go up to $10^{18}$ n$_{eq}$/cm$^2$.  At the same time there is significant electronics component testing for a variety of applications that need much lower beam rates on the order $10^{6}$ to $10^{10}$ particles/second which requires either dedicated facilities or a beam with significant range.  Reaching high fluences presents significant materials handling challenges that require purpose built facility design and remote handling.  An increased investment in irradiation facility resources is important over the coming decade.

The current schedules for the long accelerator shutdown at Fermilab for PIP-II and LBNF work, and the LS3 shutdown at CERN for the HL-LHC largely overlap.  This leaves the potential for about a two year period in the later half of this decade where both the Fermilab and CERN test beam and irradiation facilities are simultaneously down.  The community should be conscious of this when planning R\&D programs in that time frame and other facilities should anticipate increased demand where beam capabilities are compatible. 

Test beam and irradiation facilities attract interest from beyond the HEP community.  Irradiation facilities in particular are a scarce commodity in high demand both in the broader scientific community (e.g. radiation hardness testing for NASA missions) and in the private sector.  Many user facilities have a charge model for beam time, there are some such as the Fermilab test beams that do not at present and are geared toward the HEP community.  The non-HEP and private sector users present a potential revenue stream to subsidize facility operation costs while supporting the HEP mission.

\section{Survey of Existing Facilities and Capabilities}
\subsection{Test Beam Facilities}

A long list of unique facilities and capabilities that enable advancements in detector technologies exist to date. Various user facilities and centers at national labs and universities are already being used by the HEP instrumentation community. The field needs specialized, dedicated facilities for their detector development. It is critical that these core facilities continue to be supported and that new capabilities required for testing and evaluation of detectors for future experiments be created. For test beam and irradiation facilities beam particle type, energy, intensity and time profile are among the distinguishing characteristics. Furthermore, beam instrumentation and other services available to users are critical, as is the possibility for regular, affordable access.

Test beams with electrons can be divided into primary (from electron beams) and secondary beams (from hadron beams) and different energy ranges for different applications. Photon beams can be achieved from inverse Compton scattering or synchrotron radiation. Proton beams can be divided into high energy, low divergence, or low energy, high rate needs. Secondary hadron beams usually consist of a mix of pions, Kaons, etc. Muon beams are achieved by collimating the hadrons from secondary beams away from the target. Test beams with ions can be useful for cosmic ray detectors (balloon or satellite born), and for detectors used in heavy ion programs. Neutrino detectors are in need of test beams with very low energies, such as can be found in tertiary beam lines. 

Most beam lines provide instrumentation for the users, such as trigger counters, beam profile monitors, TOF, Cerenkov detectors, wire chambers, silicon telescopes, magnets. There is an effort towards the development of common tools, such as DAQ systems. Providers and users get together once a year for the workshop series on Beam Telescopes and Test Beams~\cite{BTTB} to discuss facility needs, DAQ, beam instrumentation, measurement results, hands-on sessions and tutorials.

Available test beam facilities are listed in Table~\ref{tab:testbeam_facilities}, for which the input was taken among other sources from the test beam database maintained at CERN~\cite{CERN_DB}. A few of these facilities such as at DESY or CERN have undergone recent refurbishments and upgrades, while others, e.g. FTBF at Fermilab, are in need of upgrades or new facilities to meet the demands of the future.

\begin{table}
\centering % Center table
\begin{tabular}{p{1.1in} | p{0.4in} | p{1.6in} | p{1.1in} |  p{0.9in}}
Facility  & \# of Beams & Particles       & Energy Range & Availability  \\ \hline

CERN/PS   & 2  & e,h,$\mu$(sec.) & 0.5-10~GeV/c & 9 mos/yr \\  \hline
CERN/SPS   & 4  & p(prim.); e,h,$\mu$(sec.); e,h(tert.); Pb(prim); other ion species & 20-400~GeV/c proton equivalent & 9 mos/yr \\ \hline
CERN/CLEAR & 1 & e$^{-}$ & 50-250~MeV/c & 8-9 mos/yr \\ \hline
DAFNE Frascati & 1 & e$^{+}$/e$^{-}$(prim. and sec.); photons & 25-750~MeV/c & 25-35 wks/yr \\ \hline
DESY (Hamburg, GE) & 3 & e$^+$/e$^-$(sec.); e$^-$(prim., planned) & 1-6 GeV/c; 6.3 GeV/c & 11 mos/yr \\ \hline
ELPH (Sendai, JP) & 2 & photons (tagged); e$^+$,e$^-$ (conversion) & 0.7-1.2 GeV/c; 0.1-1 GeV/c & 2 mos/yr \\ \hline
ELSA (Bonn, GE) & 1 & e$^-$ & 1.2-3.2 GeV/c & $\sim$~30 days/yr \\ \hline
FTBF (Fermilab, US) & 2 & p (prim.); e,h,$\mu$ (sec.); h (ter.) & 120 GeV/c; 1-66 GeV/c; 200-500 MeV/c & 8 mos/yr \\ \hline
IHEP (Beijing, CN) & 2 & e (prim.); e,p,$\pi$ (sec.) & 1.1-2.5 GeV/c; 0.1-1.2 GeV/c & 3 mos/yr \\ \hline
IHEP (Protvino, RU) & 5 & p,C-12 (prim.); p,K,$\pi$,$\mu$,e (sec.) & 70 (6-300) GeV/c; 1-45 GeV/c & 2 mos/yr \\ \hline
MAMI (Mainz, GE) & 3 & e$^-$,photons & $<1.6$ GeV/c & $\sim$~30 days/yr \\ \hline
NSRL (Brookhaven, US) & 1 & p, heavy ions & 0-1 GeV/n & 10 mos/yr \\ \hline
piEI,ppiMI,etc. (PSI, CH) & 2-4 & $\pi$,$\mu$,e,p & 50-450 MeV/c & 6-8 mos/yr \\ \hline
PIF (PSI, CH) & 1 & p & 5-230 MeV/c & 11 mos/yr \\ \hline
RCNP (Osaka, JP) & 7 & p,heavy ions,n,$\mu^+$ & 24-400 MeV/c & 7-8 mos/yr \\ \hline
SLAC (Stanford, US) & 0 & e (prim.); e (sec.) & 2.5-15 GeV/c; 1-14 GeV/c & currently no beam \\ \hline
SPRING-8 (Compton Facility, JP) & 2 & photons (tagged), e$^+$,e$^-$ (conv.) & 0.4-2.9 GeV/c & $>$60 days/yr \\ \hline 
\end{tabular}
\caption{Overview of existing test beam facilities}
\label{tab:testbeam_facilities}
\end{table}

\subsection{Irradiation Facilities}

Irradiation facilities are needed to develop radiation-hard detectors for HEP experiments. We use them to study radiation damage to materials, such as cables and glues, sensors, such as semiconductors, gas and crystals, as well as electronics components, such as COTS as well as ASICs. Components for accelerators as well as targetry typically face even harsher environments and have needs for more intense beam doses and fluences for testing. Different applications call for different beam conditions, rates, fluences or doses. Devices  might need to be cooled, powered and read out to look for performance degradation or operational functionality such as Single Event Effects. 

There are many different types of facilities in use to date, such as high-energy accelerators, cyclotrons, nuclear reactors or table-top sources. Access to these facilities has at times been difficult and not enough time slots for all potential users were available. We think that overall more facilities are needed in general, and in particular those that will address high-rate, high-intensity needs for future collider detectors, where the potential fluence encountered can go up two orders of magnitude compared to today's highest detector requirements. Similar to the test beam facilities, CERN also maintains a central database of irradiation facilities around the world. The list is too long to replicate here, but can be found and filtered on in~\cite{CERN_IFDB}. Instead we provide a short selection of facilities that are especially relevant to US-based efforts in Tab.~\ref{tab:irrad_facilities}.

\begin{table}
\centering % Center table
\begin{tabular}{p{1.8in} | p{0.7in} | p{1.2in} | p{0.8in} }

Facility  & Particles  & Energy Range & SEE Tests  \\ \hline

%BLIP (BNL, US) & p, n & 65-200 MeV & no \\ \hline
BASE & p, n, heavy ions & 1-55 MeV & yes \\ \hline
CERN (Geneva, CH) & p & 24 GeV & yes  \\  \hline
CNL (UC Davis, US) & p, n & 1-67.5 MeV & no \\ \hline
CYCLONE (Louvain, BE) & p, heavy ions & 14.4-65 MeV & yes?  \\ \hline
%FSU (Florida, US) & p & 17 MeV &  \\ \hline
ITA (FNAL, US) & p & 400 MeV & yes  \\ \hline
KAZ (Karlsruhe, GE) & p & ~23 MeV & yes?  \\ \hline
LANSCE (LANL, US) & p & 800 MeV & no  \\ \hline
%LEAF (ANL, US) & e, n & 55 MeV, 0.5 MeV/c &  \\\ %\hline
MC40 (Birmingham, UK) & p & 37 MeV & yes \\ \hline
McClellan (UC Davis, US) & n & 0.025 eV - 1 MeV & yes \\ \hline
PIF (TRIUMF, CA) & p & 5-500 MeV & no  \\ \hline
RINSC (Rhode Island, US) & n & - & no \\ \hline
TRIGA (Ljubljana, SI) & n & few MeV & no  \\ \hline 

\end{tabular}
\caption{Overview of some existing irradiation facilities}
\label{tab:irrad_facilities}
\end{table}

\section{Snowmass LOIs and Coverage of Needs}
In this section we summarize details of some LOIs that were submitted to Snowmass, proposing new facilities to be developed in response to current and future gaps between available facilities and user needs.

\subsection{High-intensity proton irradiation facility}
Fermilab plans to upgrade its accelerator infrastructure to deliver a 2.4 MW 120 GeV proton beam for DUNE. The Booster Replacement (BR) will replace the current 8 GeV booster with a new accelerating facility. Technology options for this upgrade include an extension of the PIP2 superconducting Linac, a rapid cycling synchrotron (RCS) as well as combinations thereof. High intensity, lower energy proton beams from initial acceleration stages, PIP2 and the BR, will be available for other experiments. The potential exists to produce lepton beams as well. Fermilab is engaging the community to explore the physics potential enabled by PIP2 and the BR, and to inform technology choices and to maximize science output. The goal is to develop these various new facilities for test beams and detector and targetry irradiation in a homogeneous and complementary way together.

The goal is to create a high-intensity proton irradiation facility at the BR or PIP2 to benefit future collider detector developments. The current Fermilab Irradiation Test Area (ITA) located at the end of the LINAC, is designed for fluences needed for the HL-LHC detector
upgrades. However, for future collider detectors doses up to two orders of magnitude higher are expected. It is paramount that detector elements under development can be tested for radiation hardness to these levels. Currently, there is no facility in the world, that would allow such tests at a reasonable timescale. It is desirable to reach on the order of $10^{18}$ protons within a few hours or days. The exact beam energy is less relevant as long as it is stable and well known. The Fermilab BR seems like an ideal candidate to serve such a facility with its high intensity proton beam. The proposal is to build a tangential arm to set up an experimental hall where devices under test (DUT) can be placed for irradiation under controlled conditions. The beam size at the DUT should be tunable and on the order of a few millimeters to five centimeters. A pulsed beam might be preferable if not necessary for cooling and readout reasons; a continuous beam would likely overheat the DUT and make cooling very challenging. Although this needs some R\&D to determine exactly which beam structure would be preferable at these high rates. Even with a pulsed beam, cooling of the DUT will be one of the main challenges. A similar time structure to the current LINAC could be suitable. The main particle type that is needed would be protons, but being able to switch to electrons, or even ions at times, would be beneficial. It would be good to have an adjacent, shielded counting room, where users could sit and operate and monitor their DUTs. This would require some user infrastructure, such as cables, power supplies, or cooling, to be placed either in the experimental hall or the adjacent counting room. Cables would be run between the two. There should be a cold, dark box for the DUTs available that can be moved in and out of the beam, including the ability to perform beam scans. Beam monitoring data need to be made available to the users as well. The irradiated DUTs would be handled by radiation safety technicians or a robot and stored cold until they can be retrieved by the
users. 

Likely, cooling of the DUTs is the main challenge and some R\&D needs to go into finding a
solution. Similarly, having robotic control of handling the box that contains the DUTs and that needs to move in and out of the beam would be highly desirable. This could be either
addressed by a commercial solution or by in-house development. The timescale by which such
a facility is needed would realistically be somewhere in the late 2020s. Before then, detector components aimed to sustain these kinds of radiation levels will likely not be ready for testing.

\subsection{New Fermilab Test Beam Facility}

The Fermilab Test Beam Facility~\cite{FTBF} (FTBF) has been operating since 2005 serving more than 200 users and 20 experiments per year covering a broad range of physics.  FTBF contains two beam lines and a variety of experimental spaces.  The beam lines deliver 120 GeV primary protons and secondary/tertiary particles down to a few hundred MeV that are extracted from the Main Injector down the Switchyard beam line.  FTBF is located at the end of over a mile of beam line coming off the Main Injector which presents increasing maintenance issues and cost with no physics benefit. The beam capabilities do not match all user requests, particularly clean low energy beam of a variety of particle species.  The program has significant need for a new purpose-built facility and beam lines in order to keep up with the demand of its user community in the coming decades.  It would be natural to jointly locate this facility in a complex with the irradiation area proposed in the previous section.

At least two beam lines should be available in the Test Beam area, but ideally 4-6 which would allow lines to deliver dedicated energies/particles without the need to insert targets.  These lines should be capable of operating simultaneously and at independent energies and intensities.  It is important at any energy that the beam focus to a manageable size, typically several centimeters or less, and have understood backgrounds and particle composition. 
120 GeV proton beam remains highly desirable for many collider R\&D efforts and tracking technologies.  There should also be beam lines capable of lower energy mixed pions and leptons from 65 GeV down to a few hundred MeV.  The most important feature here is that the beam composition be well understood, low background, controllable, and that sufficient particle identification tools exist.  The new location would benefit from extracting dedicated low energy beam from the PIP-II (800 MeV) and Booster Replacement (8 GeV) stages.  It is desirable to have at least one beamline dedicated to low energy muons which are frequently requested and not readily achievable with the existing facility. The opportunities are well documented in the 2013 Snowmass Report~\cite{Kronfeld:2013uoa}. 

The experimental area should have space for dedicated staff and staging equipment as well as electronics and mechanical work spaces.  There should be a control room for monitoring by users. The facility should have multiple beam enclosures capable of supporting 4-6 experiments simultaneously.  The facility should be instrumented to support beam monitoring, particle tracking and identification, gas systems, calorimetry, and motion tables.  There should be 1-2 tracking telescopes available to meet the demand for sensor R\&D and they should be compatible with DAQs used internationally.  All beam line instrumentation should be implemented in a universal DAQ system that is accessible to users in all experimental spaces.  At least one experimental space should have a high ceiling and be accessible by crane.  There should be space to support experiments both on the scale of weeks and months.  The floor plan should have the flexibility to accommodate installations of varying size from silicon chips up to several ton detectors. Dedicated space should support both tracking studies (i.e. testing of sensors for the HL-LHC) and calorimetry.  There should be video conferencing spaces available. In addition, magnets should be available for users. There are use cases for spectrometer magnets and for a large bore magnet that devices can be placed inside of. Cryogenic cooling would benefit noble liquid detector R\&D and is extremely useful to the community. 

\subsection{LESA CW Electron Test Beam at SLAC}
After many successful years of Test Beam in End Station A at SLAC\cite{ESTB}, the End Station Test Beam (ESTB) kickers were disconnected for construction of the LCLS-II superconducting beam line, ending the delivery of 2-15 GeV electron beams from the normal-conducting LCLS linac to End Station A (ESA) for test beam use. 

A new multi-purpose beamline, The SLAC Linac to End Station A (LESA) will extract low current near-CW 4-8 GeV beams from the SLAC LCLS-II superconducting linac and transport this beam to ESA. LESA is anticipated to be available for 250 days/year, with test-beam use for a few months per year and the remainder devoted to longer-running experiments focused on dark matter and electronuclear scattering.  The S30XL kicker and beamline, a predecessor of LESA, is currently under construction; LESA is expected to be fully operational (including the source laser) in FY24.  

As a test beam, LESA will build on the experience of ESTB and support many of the same capabilities (albeit at  lower beam energy).  The capability to reduce the LESA beam to one/few electrons at a time, with control over the beam spot, will use the same systems demonstrated in ESTB program.  
In addition, LESA offers several attractive new features as a test-beam facility. LESA can deliver a high repetition rate of up to one pulse every 21 ns, at 50\% duty cycle (specifically, a $\sim 500$ ns macro-pulse  delivered every 1.1 $\mu$s). This repetition rate can be flexibly reduced, while maintaining a well-defined timestructure, by adjusting which bunches are populated within each macro-pulse (source laser control) and/or reducing frequency of macro-pulses delivered to LESA (kicker timing). This allows for rapid accumulation of test-beam statistics --- LESA can deliver up to $10^7$ one-electron bunches per second. Combined with controllable current and timing (e.g. the Poisson mean electron multiplicity can be programmed bunch-to-bunch within a $\sim 500$ ns bunch train), LESA furnishes flexible opportunities to study out-of-time pile-up and high-rate detector performance in a controlled test-beam environment.  LESA will also deliver pulses of ps or sub-ps length. Such short pulses with known inter-bunch timing provide a valuable absolute timing reference for testing fast timing detectors and potential 4D tracking devices. By measuring detector timing against an absolute reference, developers of these detectors can clearly separate the effects of intrinsic time resolution and variation in response time across a detector, greatly facilitating R\&D in this direction. The possible delivery of currents up to $25$ nA (contingent on radiation safety studies and impact on other detectors installed in ESA) would further allow material mapping on silicon detectors, studies of EM-induced radiation damage, and testing of integrating detectors. 

A dozen distinct test-beam use cases are described in \cite{S30XL_Science}. These include applications to LHC upgrades and related experiments; to basic detector R\&D; and to detector development for the PBC program at CERN and for Nuclear Physics. 
User access will be similar to the ESTB model, with brief proposals reviewed by a Scientific Committee to evaluate and prioritize based on scientific value, feasibility, and resource requirements. 

\subsection{Water Cherenkov Test Experiment}
Water Cherenkov detector remain a major part of the experimental neutrino program, with the approved Hyper-Kamiokande experiment, and proposed experiments such as THEIA and ESSnuSB.  These experiments are improving upon past water Cherenkov detectors, with new developments in photosensors and detector media, such as Gadolinum loading in water, now deployed in Super-Kamiokande, and the developing technology of water-based liquid scintillator (WbLS).  At the same time, these very large detectors will require percent-level calibration to take full advantage of the large data sets they will collect.  The long-baseline experiments will also have kiloton scale water Cherenkov near detectors, such as the Hyper-K Intermediate Water Cherenkov Detector (IWCD). 

All of these developments and challenges in water-based detectors point to the need for a prototype water Cherenkov detector that can operate in a particle test beam while deploying new detector technologies and calibration techniques.  This experiment should also have the capability to measure important physics processes for water Cherenkov detectors and neutrino interactions such as Chrenenkov light production and the scattering and absorption of hadrons in water.  The Water Cherenkov Test Experiment (WCTE) is a planned 3.5~m tall by 4~m diameter detector that will initially be instrumented with detector systems planned for the Hyper-K IWCD.  It is planned to operate in the CERN PS T9 beam line.  Since the experiment should accept particle momenta down to 200~MeV/c or less, including pions, a tertiary production target located 3~m upstream from the WCTE and a compact spectrometer are planned.  Particle identification will be done with time-of-flight and aerogel Cherenkov threshold detectors.  The WCTE will also accept the secondary beam from the T9 beamline.  Recent upgrades to the T9 beam line magnets may allow operation down to 200~MeV/c or less.  Given this development the possibility of operation without the tertiary production target is also being studied, including the capability to separate muons and pions using a combination of time-of-flight and aerogel Cherenkov threshold detectors.

The WCTE is approved for an initial 12~week run in the T9 beam line, which will focus on operation with pure-water and Gadolinium loaded water.  The WCTE experimental collaboration plans to engage with the community on the possibility of a more long-term installation of the WCTE in order to facility test beam measurements for future photosensor developments, such as LAPPDs and dichroicons, and detector media developments such as WbLS.  The more long-term operation of WCTE may also be beneficial to operating experiments if they find the need for further test beam measurements to help control systematic uncertainties.

\def\thefootnote{\fnsymbol{footnote}}
\setcounter{footnote}{0}

\bibliographystyle{IEEEtran}  
\bibliography{references}

\begin{thebibliography}{1}
\providecommand{\url}[1]{#1}
\csname url@samestyle\endcsname
\providecommand{\newblock}{\relax}
\providecommand{\bibinfo}[2]{#2}
\providecommand{\BIBentrySTDinterwordspacing}{\spaceskip=0pt\relax}
\providecommand{\BIBentryALTinterwordstretchfactor}{4}
\providecommand{\BIBentryALTinterwordspacing}{\spaceskip=\fontdimen2\font plus
\BIBentryALTinterwordstretchfactor\fontdimen3\font minus
  \fontdimen4\font\relax}
\providecommand{\BIBforeignlanguage}[2]{{%
\expandafter\ifx\csname l@#1\endcsname\relax
\typeout{** WARNING: IEEEtran.bst: No hyphenation pattern has been}%
\typeout{** loaded for the language `#1'. Using the pattern for}%
\typeout{** the default language instead.}%
\else
\language=\csname l@#1\endcsname
\fi
#2}}
\providecommand{\BIBdecl}{\relax}
\BIBdecl

\bibitem{brn2019}
\BIBentryALTinterwordspacing
B.~Fleming \emph{et~al.}, ``Basic research needs for high energy physics
  detector research \& development: Report of the office of science workshop on
  basic research needs for hep detector research and development: December
  11-14, 2019,'' 12 2019. [Online]. Available:
  \url{https://www.osti.gov/biblio/1659761}
\BIBentrySTDinterwordspacing

\bibitem{Survey}
\BIBentryALTinterwordspacing
``Future test beam facility survey.'' [Online]. Available:
  \url{https://forms.gle/TUwG6b6EYbCtgmtz7}
\BIBentrySTDinterwordspacing

\bibitem{BTTB}
\BIBentryALTinterwordspacing
``Beam telescopes and test beams workshop series.'' [Online]. Available:
  \url{https://indico.cern.ch/event/1058977/}
\BIBentrySTDinterwordspacing

\bibitem{CERN_IFDB}
\BIBentryALTinterwordspacing
``Cern irradiation facilities database.'' [Online]. Available:
  \url{https://irradiation-facilities.web.cern.ch/publicDB.php}
\BIBentrySTDinterwordspacing

\bibitem{FTBF}
\BIBentryALTinterwordspacing
``Ftbf website.'' [Online]. Available: \url{https://ftbf.fnal.gov}
\BIBentrySTDinterwordspacing

\bibitem{Kronfeld:2013uoa}
U.~Al-Binni \emph{et~al.}, ``{Project X: Physics Opportunities},'' 6 2013.

\bibitem{ESTB}
{Pivi, Mauro (Apr 2011). ``ESTB: A New Beam Test Facility at SLAC''
  (SLAC-PUB-14424). United States}.

\bibitem{S30XL_Science}
{A description of LESA and its science program can be found in SLAC-R-1147.The
  LESA beamline has been known as DASEL, S30XL, and YABBL.}

\end{thebibliography}

\end{document}